\pdfoutput=1
\documentclass[12pt,english]{aastex}
\usepackage[T1]{fontenc}
\usepackage[latin9]{inputenc}
\usepackage{array}
\usepackage{mathtools}
\usepackage{amsbsy}
\usepackage{amssymb}
\usepackage{cancel}
\usepackage{mathdots}
\usepackage{stmaryrd}
\usepackage{stackrel}
\usepackage{graphicx}
\PassOptionsToPackage{version=3}{mhchem}
\usepackage{mhchem}
\usepackage{esint}

\makeatletter

\providecommand{\tabularnewline}{\\}

\usepackage{amsbsy}

\usepackage{babel}

\usepackage{babel}

\makeatother

\usepackage{babel}
\begin{document}

\title{Effects of large-scale non-axisymmetric perturbations in the mean-field
solar dynamo.}

\author{V.V. Pipin$^{1-4}$ and A.G. Kosovichev$^{3,4,5}$}

\affil{$^{1}$Institute of Solar-Terrestrial Physics, Russian Academy of
Sciences, \\
 $^{2}$ Institute of Geophysics and Planetary Physics, UCLA, Los
Angeles, CA 90065, USA \\
 $^{3}$W.W. Hansen Experimental Physics Laboratory, Stanford University,
Stanford, CA 94305, USA \\
 $^{4}$NASA Ames Research Center, Moffett Field, CA 94035, USA \\
 $^{5}$New Jersey Institute of Technology, CA 92314, USA }
\begin{abstract}
We explore a response of a non-linear non-axisymmetric mean-field
solar dynamo model to shallow non-axisymmetric perturbations. After
a relaxation period the amplitude of the non-axisymmetric field depends
on the initial condition, helicity conservation, and the depth of
perturbation. It is found that a perturbation which is anchored at
$0.9R_{\odot}$ has a profound effect on the dynamo process, producing
a transient magnetic cycle of the axisymmetric magnetic field, if
it is initiated at the growing phase of the cycle. The non-symmetric
with respect to the equator perturbation results in a hemispheric
asymmetry of the magnetic activity. The evolution of the axisymmetric
and non-axisymmetric fields depend on the turbulent magnetic Reynolds
number $R_{m}$. In the range of $R_{m}=10^{4}-10^{6}$ the evolution
returns to the normal course in the next cycle, in which the non-axisymmetric
field is generated due to a non-linear $\alpha$-effect and magnetic
buoyancy. In the stationary state the large-scale magnetic field demonstrates
a phenomenon of ``active longitudes'' with cyclic 180$^{\circ}$
``flip-flop'' changes of the large-scale magnetic field orientation.
The flip-flop effect is known from observations of solar and stellar
magnetic cycles. However, this effect disappears in the model which
includes the meridional circulation pattern determined by helioseismology.
The rotation rate of the non-axisymmetric field components varies
during the relaxation period, and carries important information about
the dynamo process. 
\end{abstract}

\section{Introduction}

Dynamo theories commonly assume that the magnetic activity of the
Sun is approximately axisymmetric on large spatial (size of the Sun)
and temporal (the period of solar cycle) scales. These models provide
a quantitative self-consistent description of the 22-year solar magnetic
cycles, and allow us to investigate the basic mechanisms of the solar
dynamo. However, deviations from the axisymmetry are rather strong
at any particular moment of observations. Intermittent patterns of
magnetic fields on the solar surface are formed because the magnetic
field emerges on the surface like separated magnetic patches, e.g,
in the form of sunspot groups. Such phenomena make a significant contribution
to the large-scale non-axisymmetric magnetic field of the Sun. 

{\citet{rad86AN} discussed dynamo generation of large-scale
non-axisymmetric (NA) magnetic field on the Sun. It was found that
the differential rotation suppresses generation of the NA magnetic
field. Non-linear dynamo processes can maintain a weak non-axisymmetric
field in expense of the axisymmetric (AS) magnetic field (see, \citealt{radler90,moss99,el2005}).
Also it was found that the dynamo generated NA magnetic field rotates
rigidly. Further theoretical developments (see, e.g., \citealt{moss91,ruz04,moss04,berd06})
showed that some of the properties of the non-linear non-axisymmetric
mean-field dynamo can be invoked for interpretation of the origin
and evolution of the so-called ``active-longitudes'' of solar magnetic
activity.  Phenomenon of the active longitudes (AL) is probably one
of the most interesting manifestations of the solar non-axisymmetric
magnetic field. It appears when the solar activity persists within
the fixed interval of longitudes for a long period of time (see e.g.,
\citealt{vitinsk,1969SoPh....7...28B,vetal86}).}

{We have to mention that the question about persistence of
the AL on century time scales interval remains a highly controversial
issue both from the observational and theoretical points of view.
For example, \citet{ber2004}, \citet{berd06} and \citet{zh2011}
reported about the AL which are persistent over a century long time
interval. However, \citet{pelt10} found that the AL have the maximal
lifetime of about one solar cycle. Another phenomenon which is related
to the AL is the so-called ``hot spots'' of the solar flare activity
\citep{bai1987,2003ApJ...585.1114B}. The reason for the different
name is because in general the longitudinal position of the activity
nests is different in the Northern and Southern hemispheres. Bai (2003)
found that those ``hot spot could persist rigidly rotating with period
about 27 days up to three solar cycles. The nonlinear dynamo models
predict that the energy of the non-axisymmetric modes is only about
$10^{-4}$ of the energy of the axisymmetric (AS) component \citep{berd06}.
It is not clear how such a weak magnetic field can modulate the nests
of the sunspots activity.}

{While, the standard mean-field model can not explain the origin
of the NA magnetic field \citep{rad86AN}, it was noted that the solar
activity produces the large-scale NA modes which are well seen in
the coronal hole configurations \citep{stix77}. The very recent example
of such events was observed by the SDO/HMI during the last decade
of May 2015. }Observations at the Wilcox Solar Observatory \citep{WSO1,WSO2}
found that the strength of the NA modes of the radial magnetic field,
e.g., the mode with the azimuthal number m=1, can be about 1 G during
epoch of the solar maxima. The axisymmetric dipole has the same magnitude
during the solar minims. It is likely that the origin of this NA field
is related to decay of solar active regions. However it is unclear
how the evolution of such NA field may impact the solar dynamo process.
The effect of the NA field on the global dynamo has not been studied
before.

{In this paper we explore a non-linear response of a mean-field
dynamo model to a shallow NA m=1 perturbations with the field strength
of 1G. It is assumed that these perturbations result from abrupt instability-type
events, which can be described by injecting NA perturbations into
the system over a very short period of time.} We consider a fairly
complete theoretical description of the mean turbulent electro-motive
force taking into account the known properties of the solar convection
zone and including the anisotropic turbulent effects due to the global
rotation. The model includes a nonlinear magnetic buoyancy effect,
and two types of non-linearity in the $\alpha$-effect, described
as ``algebraic'' and ``dynamical'' quenching \citep{twork}. The
algebraic quenching is due to the back-reaction of the dynamo-generated
magnetic field on helical turbulence. The dynamical quenching results
from a magnetic helicity conservation condition \citep{kleruz82}.
We will show that the nonlinear non-axisymmetric effects are sufficiently
strong to reproduce the ``flip-flop'' phenomenon and explain the
rotation rate of active longitudes on the Sun \citep{tbk02,berd06,gy12}.

\section{Basic equations}

Evolution of the large-scale magnetic field in perfectly conductive
media is described by the mean-field induction equation \citep{KR80,moff:78,park}:
\begin{equation}
\partial_{t}\left\langle \mathbf{B}\right\rangle =\boldsymbol{\nabla}\times\left(\mathbf{\boldsymbol{\boldsymbol{\mathcal{E}}}+}\left\langle \mathbf{U}\right\rangle \times\left\langle \mathbf{B}\right\rangle \right)\label{eq:mfe}
\end{equation}
where $\boldsymbol{\mathcal{E}}=\left\langle \mathbf{u\times b}\right\rangle $
is the mean electromotive force; $\mathbf{u}$ and $\mathbf{b}$ are
the turbulent fluctuating velocity and magnetic field respectively;
and $\left\langle \mathbf{U}\right\rangle $ and $\left\langle \mathbf{B}\right\rangle $
are the mean velocity and magnetic field. Our solution of the dynamo
equation will follow the outline given earlier by \citet{moss91}
and \citet{moss99}. For convenience we decompose the magnetic field
into the axisymmetric (AS), (hereafter $\overline{\mathbf{B}}$-field),
and non-axisymmetric (NA) parts, (hereafter $\tilde{\mathbf{B}}$-field):
$\left\langle \mathbf{B}\right\rangle =\overline{\mathbf{B}}+\tilde{\mathbf{B}}$.
We assume that the mean flow is axisymmetric $\left\langle \mathbf{U}\right\rangle \equiv\overline{\mathbf{U}}$.
Let $\hat{\boldsymbol{\phi}}=\mathbf{e_{\phi}}$ and $\hat{\mathbf{r}}=r\mathbf{e}_{r}$
be vectors in the azimuthal and radial directions respectively, then
we represent the mean magnetic field vectors as follows: 
\begin{eqnarray}
\left\langle \mathbf{B}\right\rangle  & = & \overline{\mathbf{B}}+\tilde{\mathbf{B}}\label{eq:b0}\\
\mathbf{\overline{B}} & = & \hat{\boldsymbol{\phi}}B+\nabla\times\left(A\hat{\boldsymbol{\phi}}\right)\label{eq:b1}\\
\tilde{\mathbf{B}} & = & \boldsymbol{\nabla}\times\left(\hat{\mathbf{r}}T\right)+\boldsymbol{\nabla}\times\boldsymbol{\nabla}\times\left(\hat{\mathbf{r}}S\right),\label{eq:b2}
\end{eqnarray}
where $A$, $B$, $T$ and $S$ are scalar functions representing
the AS and NA parts respectively. Assuming that $A$ and $B$ do not
depend on longitude, Eqs(\ref{eq:b1}, \ref{eq:b2}) ensure that the
field $\left\langle \mathbf{B}\right\rangle $ is divergence-free.
Taking the scalar product of Eq(\ref{eq:mfe}) with vector $\hat{\boldsymbol{\phi}}$
we get equations for the AS magnetic field components, 
\begin{align}
\partial_{t}B & =\hat{\boldsymbol{\phi}}\cdot\boldsymbol{\nabla}\times\left(\mathbf{\boldsymbol{\boldsymbol{\mathcal{E}}}+}\mathbf{\overline{U}}\times\overline{\mathbf{B}}\right),\label{eq:bf}\\
\partial_{t}A & =\hat{\boldsymbol{\phi}}\cdot\left(\mathbf{\boldsymbol{\mathcal{E}}+}\overline{\mathbf{U}}\times\mathbf{\overline{B}}\right),\label{eq:af}
\end{align}
To get equation for $T$ we take curl of Eq(\ref{eq:mfe}), and then
calculate the scalar product with vector $\mathbf{\hat{r}}$. Similarly,
equation for $S$ is obtained by taking twice curl of Eq(\ref{eq:mfe})
and then the scalar product with vector $\mathbf{\hat{r}}$. The procedure
is described in detail by \citet{KR80}. Equations for the NA field
are

\begin{eqnarray}
\partial_{t}\Delta_{\Omega}T & = & \Delta_{\Omega}V^{(U)}+\Delta_{\Omega}V^{(\mathcal{E})},\label{eq:T}\\
\partial_{t}\Delta_{\Omega}S & = & \Delta_{\Omega}U^{(U)}+\Delta_{\Omega}U^{(\mathcal{E})},\label{eq:S}
\end{eqnarray}
where ${\displaystyle \Delta_{\Omega}=\frac{\partial}{\partial\mu}\sin^{2}\theta\frac{\partial}{\partial\mu}+\frac{1}{\sin^{2}\theta}\frac{\partial^{2}}{\partial\phi^{2}}}$,
$\mu=\cos\theta$ and $\theta$ is a polar angle, and 
\begin{eqnarray}
\Delta_{\Omega}V^{(U)} & = & -\hat{\mathbf{r}}\cdot\boldsymbol{\nabla}\times\boldsymbol{\nabla}\times\left(\mathbf{\overline{U}}\times\mathbf{\tilde{\mathbf{B}}}\right),\label{eq:vu}\\
\Delta_{\Omega}V^{(\mathcal{E})} & = & -\hat{\mathbf{r}}\cdot\boldsymbol{\nabla}\times\boldsymbol{\nabla}\times\boldsymbol{\boldsymbol{\mathcal{E}}},\label{eq:ve}\\
\Delta_{\Omega}U^{(U)} & = & -\hat{\mathbf{r}}\cdot\boldsymbol{\nabla}\times\left(\mathbf{\overline{U}}\times\tilde{\mathbf{B}}\right),\label{eq:uu}\\
\Delta_{\Omega}U^{(\mathcal{E})} & = & -\hat{\mathbf{r}}\cdot\boldsymbol{\nabla}\times\boldsymbol{\boldsymbol{\mathcal{E}}}.\label{eq:ue}
\end{eqnarray}
The scalar functions with superscript $^{(U)}$ contain contributions
from the large-scale AS flows like the differential rotation or meridional
circulation. The integration domain includes the solar convection
zone from $0.71$ to $0.99R_{\odot}$. The distribution of the mean
flows is given by helioseismology (\citealt{Howe2011JPh} and \citealt{Zhao13m}).
Profiles of the angular velocity and meridional circulation are illustrated
in Figure 1. 
\begin{figure}
\includegraphics[width=0.7\textwidth]{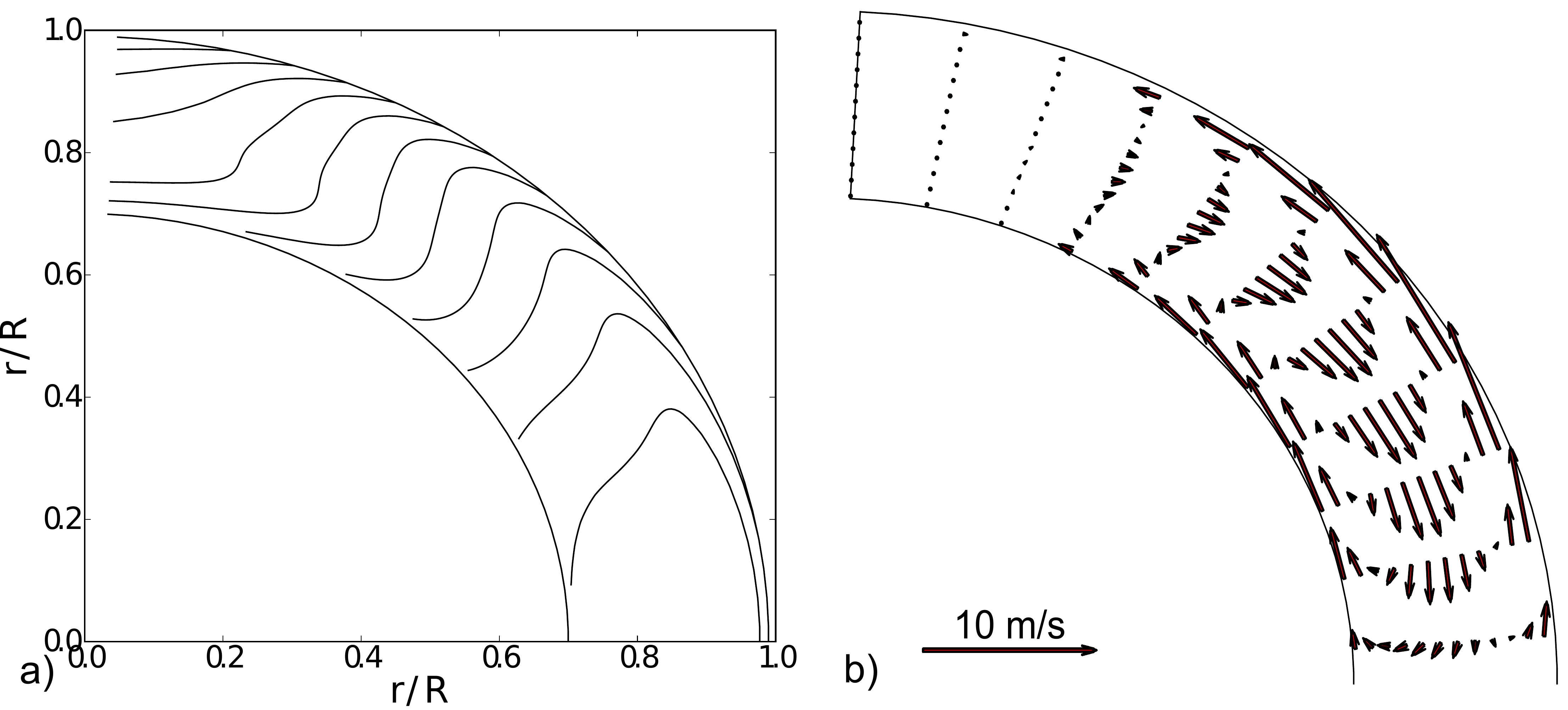}\protect\protect \protect\caption{\label{one}a) The isolines of constant angular velocity ranging from
0.6$\Omega_{0}$ to 0.96$\Omega_{0}$ ($\Omega_{0}=2.87\times10^{-6}$s$^{-1}$);
b) illustration of the double-cell meridional circulation model, consistent
with the helioseismology results of \citet{Zhao13m}.}
\end{figure}

We use formulation for the mean electromotive force obtained by Pipin(2008).
The calculations of the mean electromotive force are done using the
mean-field magnetohydrodynamics framework and the so-called ``minimal
tau-approximation'' (see, e.g., \citealp{bf:02,rad-kle-rog,2005PhR...417....1B}).
The tau-approximation suggests that the second-order correlations
do not vary significantly on timescale $\tau_{c}$ which corresponds
to a typical turnover time of the convective flows. The theoretical
calculations are performed for anelastic turbulent flows. They take
into account the effects of density stratification, spatial inhomogeneity
of the intensity of turbulent flows, and inhomogeneity of the large-scale
magnetic fields. The effects of large-scale inhomogeneity of the turbulent
flows and magnetic fields are computed in the first order of the Taylor
expansion in terms of ratio $\ell/L$, where $\ell$ is a typical
spatial scale of the turbulence, and $L$ is a spatial scale of the
mean quantities. The mean electromotive force, $\boldsymbol{\mathcal{E}}$,
is expressed as follows (Pipin, 2008): 
\begin{equation}
\mathcal{E}_{i}=\left(\alpha_{ij}+\gamma_{ij}\right)\left\langle B\right\rangle _{j}-\eta_{ijk}\nabla_{j}\left\langle B\right\rangle _{k}.\label{eq:EMF-1}
\end{equation}
where symmetric tensor $\alpha_{ij}$ models the generation of magnetic
field by the $\alpha$- effect; antisymmetric tensor{ }$\gamma_{ij}$
controls the mean drift of the large-scale magnetic fields in turbulent
medium; tensor $\eta_{ijk}$ governs the turbulent diffusion. We take
into account the effect of rotation and magnetic field on the mean-electromotive
force (see, e.g., \citealt{pi15M} for details). To determine unique
solution of Eqs.(\ref{eq:bf}-\ref{eq:S}) we apply the following
gauge{ }(see, e.g., \citealt{KR80,ruz04}): 
\begin{equation}
\int_{0}^{2\pi}\int_{-1}^{1}Sd\mu d\phi=0,\,\,\int_{0}^{2\pi}\int_{-1}^{1}Td\mu d\phi=0.\label{eq:norm}
\end{equation}
The same gauge is used in Eqs(\ref{eq:vu}-\ref{eq:ue}).

\subsection{Nonlinear interaction of the axisymmetric and non-axisymmetric modes}

Interaction between the AS and NA modes in the mean-field dynamo models
can be due to nonlinear dynamo effects, for example, the $\alpha$-effect,
\citep{KR80,moss99}. In our model the $\alpha$ effect takes into
account the kinetic and magnetic helicities in the following form:
\begin{eqnarray}
\alpha_{ij} & = & C_{\alpha}\sin^{2}\theta\psi_{\alpha}(\beta)\alpha_{ij}^{(H)}\eta_{T}+\alpha_{ij}^{(M)}\frac{\left\langle \chi\right\rangle \tau_{c}}{4\pi\overline{\rho}\ell^{2}}\label{alp2d}
\end{eqnarray}
where $C_{\alpha}$ is a free parameter which controls the strength
of the $\alpha$- effect due to turbulent kinetic helicity; $\alpha_{ij}^{(H)}$
and $\alpha_{ij}^{(M)}$ express the kinetic and magnetic helicity
parts of the $\alpha$-effect, respectively; $\eta_{T}$ is the magnetic
diffusion coefficient, and $\left\langle \chi\right\rangle =\left\langle \mathbf{a}\cdot\mathbf{b}\right\rangle $
($\mathbf{a}$ and $\mathbf{b}$ are the fluctuating parts of magnetic
field vector-potential and magnetic field vector). Both the $\alpha_{ij}^{(H)}$
and $\alpha_{ij}^{(M)}$ depend on the Coriolis number $\Omega^{*}=4\pi{\displaystyle \frac{\tau_{c}}{P_{rot}}}$,
where $P_{rot}$ is the rotational period, $\tau_{c}$ is the convective
turnover time, and $\ell$ is a typical length of the convective flows
(the mixing length). A theoretical justification for the latitudinal
factor, $\sin^{2}\theta$, in Eq(\ref{alp2d}) was given by \citet{2003PhRvE..67b6321K}.
Function $\psi_{\alpha}(\beta)$ controls the so-called ``algebraic''
quenching of the $\alpha$- effect where $\beta=\left|\mathbf{\left\langle B\right\rangle }\right|/\sqrt{4\pi\overline{\rho}u'^{2}}$,
$u'$ is the RMS of the convective velocity. For the case of the strong
magnetic field, $\beta\gg$1, $\psi_{\alpha}\sim\beta^{-2}$ . The
interaction between the axisymmetric and non-axisymmetric dynamo modes
via $\psi_{\alpha}(\beta)$ is because both modes contribute to parameter
$\beta$. Also, for the case $\beta>$1, the latitudinal profile of
the $\alpha$ effect changes. This can affect the dynamo conditions
for excitation of the NA modes. \citet{radler90} and \citet{moss99}
discussed the evolution of NA magnetic field in a simple dynamo model
with such ``algebraic'' quenching.

The dynamical quenching is caused by the magnetic helicity conservation
(see, \citealt{kleruz82}). This effect was discovered by \citet{pouquet-al:1975a}
and \citet{pouquet-al:1975b}. Contribution of the magnetic helicity
to the $\alpha$-effect is expressed by the second term in Eq.(\ref{alp2d}).
The magnetic helicity density of turbulent field, $\left\langle \chi\right\rangle =\left\langle \mathbf{a}\cdot\mathbf{b}\right\rangle $,
is governed by the conservation law \citep{hub-br12,pip13M}: 
\begin{equation}
\frac{\partial\left\langle \chi\right\rangle ^{(tot)}}{\partial t}=-\frac{\left\langle \chi\right\rangle }{R_{m}\tau_{c}}-2\eta\left\langle \mathbf{B}\right\rangle \cdot\left\langle \mathbf{J}\right\rangle -\boldsymbol{\nabla\cdot}\boldsymbol{\boldsymbol{\mathcal{F}}}^{\chi},\label{eq:helcon-1}
\end{equation}
where $\left\langle \chi\right\rangle ^{(tot)}=\left\langle \chi\right\rangle +\left\langle \mathbf{A}\right\rangle \cdot\left\langle \mathbf{B}\right\rangle $
is the total magnetic helicity density of the mean and turbulent fields,
$\boldsymbol{\boldsymbol{\mathcal{F}}}^{\chi}=-\eta_{\chi}\boldsymbol{\nabla}\left\langle \chi\right\rangle $
is the diffusive flux of the turbulent magnetic helicity, and $R_{m}$
is the magnetic Reynolds number. The coefficient of the turbulent
helicity diffusivity, $\eta_{\chi}$, is chosen ten times smaller
than the isotropic part of the magnetic diffusivity \citep{mitra10}:
$\eta_{\chi}=\frac{1}{10}\eta_{T}$. Similarly to the magnetic field,
the mean magnetic helicity density can be formally decomposed into
the axisymmetric and non-axisymmetric parts: $\left\langle \chi\right\rangle ^{(tot)}=\overline{\chi}^{(tot)}+\tilde{\chi}^{(tot)}$.
The same can be done for the magnetic helicity density of the turbulent
field: $\left\langle \chi\right\rangle =\overline{\chi}+\tilde{\chi}$,
where $\overline{\chi}=\overline{\mathbf{a}\cdot\mathbf{b}}$ and
$\tilde{\chi}=\tilde{\left\langle \mathbf{a}\cdot\mathbf{b}\right\rangle }$.
Then we have, 
\begin{align}
\overline{\chi}^{(tot)} & =\overline{\chi}+\overline{\mathbf{A}}\cdot\overline{\mathbf{B}}+\overline{\tilde{\mathbf{A}}\cdot\tilde{\mathbf{B}}},\label{eq:t1}\\
\tilde{\chi}^{(tot)} & =\tilde{\chi}+\overline{\mathbf{A}}\cdot\tilde{\mathbf{B}}+\tilde{\mathbf{A}}\cdot\overline{\mathbf{B}}+\tilde{\mathbf{A}}\cdot\tilde{\mathbf{B}},\label{eq:t2}
\end{align}
Evolution of the $\overline{\chi}$ and $\tilde{\chi}$ is governed
by the corresponding parts of Eq(\ref{eq:helcon-1}). Thus, the model
takes into account contributions of the AS and NA fields in the whole
magnetic helicity density balance, providing a non-linear coupling.
We see that the $\alpha$-effect is dynamically linked to the longitudinally
averaged magnetic helicity of the NA $\tilde{\mathbf{B}}$-field,
which is the last term in Eq(\ref{eq:t1}). Thus, the nonlinear $\alpha$-effect
is non-axisymmetric, and it results in coupling between the AS and
NA modes. The coupling works in both directions. For instance, the
azimuthal $\alpha$-effect results in $\mathcal{E}_{\phi}=\alpha_{\phi\phi}\left\langle B_{\phi}\right\rangle $.
If we denote the NA part of the $\alpha_{\phi\phi}$ by $\tilde{\alpha}_{\phi\phi}$
then the mean electromotive force is $\overline{\mathcal{E}}_{\phi}=\overline{\alpha}_{\phi\phi}\overline{B}_{\phi}+\overline{\tilde{\alpha}_{\phi\phi}\tilde{B}_{\phi}}$.
This introduces a new source in Eq(\ref{eq:af}) which is usually
ignored in the axisymmetric dynamo models.

Magnetic buoyancy is another nonlinear effect which is important in
the large-scale dynamo. The part of the mean electro-motive force
which is responsible for magnetic buoyancy is \citep{kp93}: 
\begin{equation}
\boldsymbol{\boldsymbol{\mathcal{E}}}^{(\beta)}=V_{\beta}\hat{\mathbf{r}}\times\mathbf{B},\label{eq:emfb}
\end{equation}
where $V_{\beta}={\displaystyle C_{\beta}\frac{\alpha_{MLT}u'}{\gamma}}\beta^{2}K\left(\beta\right),$
$u'$ is the RMS convection velocity, $K\left(\beta\right)$ {can
be found in }\citep{kp93,pi15M}, $\gamma$ is the adiabatic exponent,
$\alpha_{MLT}$ is the mixing-length theory parameter, $C_{\beta}$
is a free parameter to switch on/off this effect in the model. For
the case $\beta\ll1$, $K\sim1$ and the up-flow velocity, $V_{\beta}$,
is proportional to the pressure of large-scale magnetic field. Similarly
to the $\alpha$-effect, the $V_{\beta}$ is non-axisymmetric and
contributes to the source terms in Eqs(\ref{eq:bf},\ref{eq:af}).
Note that advection of the large-scale magnetic field by the magnetic
buoyancy reduces concentration of the magnetic field near the bottom
of the convection zone, and increases the field strength near the
top.

\subsection{Parameters of the convection zone and numerical procedure}

The distribution of the turbulent parameters, such as the typical
convective turn-over time, $\tau_{c}$, the mixing length, $\ell$,
and the RMS convection velocity, $u'$, are taken from the solar interior
model of \citet{stix:02}. We define the mixing-length: $\ell=\alpha_{{\rm MLT}}\left|\Lambda^{(p)}\right|^{-1}$,
where $\Lambda{}^{(p)}=\boldsymbol{\nabla}\log\overline{p}\,$ is
the inverse pressure scale, and the mixing-length parameter $\alpha_{{\rm MLT}}=2$.
The profile of the turbulent diffusivity is taken in the form $\eta_{T}=C_{\eta}{\displaystyle \frac{u'^{2}\tau_{c}}{3f_{ov}\left(r\right)}}$,
where $f_{ov}(r)=1+\exp\left[50\left(r_{ov}-r\right)\right]$, $r_{ov}=0.725R_{\odot}$
controls quenching of the turbulent effects near the bottom of the
convection zone, which is $r_{b}=0.715R$. Free parameter $C_{\eta}$,
$\left(0<C_{\eta}<1\right)$ controls the efficiency of mixing of
the large-scale magnetic field by turbulence. It is usually employed
to tune the period of the dynamo cycle.

The numerical scheme employs the spherical harmonics decomposition
for the non-axisymmetric part of the problem, i.e., the scalar functions
T and S in Eqs(\ref{eq:T},\ref{eq:S}) are represented in the form:
\begin{align}
T\left(r,\mu,\phi,t\right) & =\sum\hat{T}_{l,m}\left(r,t\right)\bar{P}_{l}^{\left|m\right|}\exp\left(im\phi\right),\label{eq:tdec}\\
S\left(r,\mu,\phi,t\right) & =\sum\hat{S}_{l,m}\left(r,t\right)\bar{P}_{l}^{\left|m\right|}\exp\left(im\phi\right),\label{eq:sdec}
\end{align}
where $\bar{P}_{l}^{m}$ is the normalized associated Legendre function
of degree $l\ge1$ and order $m\ge1$. The simulations which we will
discuss include 600 spherical harmonics ($l_{max}=28$). Note that
$\hat{S}_{l,-m}=\hat{S}_{l,m}^{*}$ and the same for $\hat{T}$. We
employ the pseudo-spectral approach for integration along latitude.
The second-order finite differences are used for discretization int
the radial direction. The numerical integration is carried out in
latitude from the pole to pole and in radius from $r_{b}=0.715R_{\odot}$
to $r_{e}=0.99R_{\odot}$. All the nonlinear terms are calculated
in the real space. The transformation between the spectral spherical
harmonic and the real 3D space was done using the Intel Fortran FFT
library. We implement algorithms of \citet{spha} to speed-up the
transform calculations. At the bottom of the convection zone we set
up a perfectly conducting boundary condition for the axisymmetric
magnetic field, and for the non-axisymmetric field we set the functions
$S$ and $T$ to zero. At the top of the convection zone the poloidal
field is smoothly matched to the external potential field. The boundary
conditions for toroidal field allow field penetrate the surface (\citet{moss91},
\citet{pk11apjl}):

\begin{eqnarray}
\delta\frac{\eta_{T}}{r_{e}}B+\left(1-\delta\right)\mathcal{E}_{\theta} & = & 0,\label{eq:tor-vac}\\
\frac{\delta}{R}T-\left(1-\delta\right)\frac{\partial T}{\partial r} & = & 0\label{eq:tb}
\end{eqnarray}
where parameter $\delta=0.99$.

The particular choice of parameters was discussed in our previous
papers (see, e.g, \citealt{2014ApJ_pipk}). The free parameters are
$C_{\alpha}=0.04$, $C_{\delta}=\frac{1}{3}C_{\alpha}$, $C_{\eta}=\frac{1}{15}$
and the anisotropy parameter $a=3$ (see \citealt{2014ApJ_pipk}).
The $\alpha$-effect parameter $C_{\alpha}$ is about 30\% above the
dynamo generation threshold. For the chosen values of $C_{\eta}$
and $a$, the turbulent diffusion coefficient in the near-surface
shear layer, at $r=0.9R_{\odot}$ is about $10^{9}$m$^{2}$s$^{-1}$
which is in agreement with surface observations (see, \citealt{abr11}).
The magnetic helicity conservation is determined by the magnetic Reynolds
number $R_{m}$, for which we considered values: $10^{4}$- $10^{6}$. 

To investigate the influence of meridional circulation we consider
two models: M1 without meridional circulation and model M2 with the
double-cell meridional circulation with a characteristic velocity
10 m/s. {For the model M2 we employ the larger parameter $C_{\alpha}$
because this model has a larger critical dynamo threshold (see, \citealt{PK13,2014ApJ_pipk}).
For this value of $C_{\alpha}$ model M2 has the same magnitude of
the generated AS toroidal magnetic field inside the convection zone
as the model M1. }The set of parameters in the models is summarized
in Table 1.

{To quantify the mirror symmetry type of the toroidal magnetic
field relative to the equator we introduce the parity index, $P$,
as follows,
\begin{eqnarray}
\overline{P} & = & \frac{\overline{E}_{S}-\overline{E}_{A}}{\overline{E}},\label{eq:par}\\
\overline{E}_{S,A} & = & \frac{1}{4}\int\left(B\left(r_{s},\theta\right)\pm B\left(r_{s},-\theta\right)\right)^{2}\sin\theta d\theta,\nonumber 
\end{eqnarray}
where ``$+$'' corresponds to the index ``S'', and ``$-$''
is for ``A''; $\overline{E}=\overline{E}_{S}+\overline{E}_{A}$,
$\overline{E}_{S}$ and $\overline{E}_{A}$ are energies of the symmetric
and anti-symmetric components of the AS toroidal magnetic field at
$r_{s}=0.9R_{\odot}$ . Similarly we define parameter $\tilde{P}$
for the NA magnetic field,
\begin{eqnarray}
\tilde{P} & = & \frac{\tilde{E}_{S}-\tilde{E}_{A}}{\tilde{E}},\label{eq:parn}\\
\tilde{E}_{S,A} & = & \int\left(\tilde{B}_{\phi}\left(r_{s},\theta,\phi\right)\pm\tilde{B}_{\phi}\left(r_{s},-\theta,\phi\right)\right)^{2}\sin\theta d\theta d\phi,\nonumber 
\end{eqnarray}
where $\tilde{E}=\tilde{E}_{S}+\tilde{E}_{A}$. The parity of the
total magnetic field is $P=\left(\overline{P}\overline{E}+\tilde{P}\tilde{E}\right)/\left(\overline{E}+\tilde{E}\right)$.}
{We will see that perturbation can affect the cycle amplitude.
Similar to \citet{radler90} we introduce a parameter to measure deviation
of the toroidal magnetic field from symmetry about the axis of rotation
\begin{equation}
M=1-\frac{\overline{E}}{\overline{E}+\tilde{E}}.\label{eq:max}
\end{equation}
}

\begin{figure}
\includegraphics[width=1\textwidth]{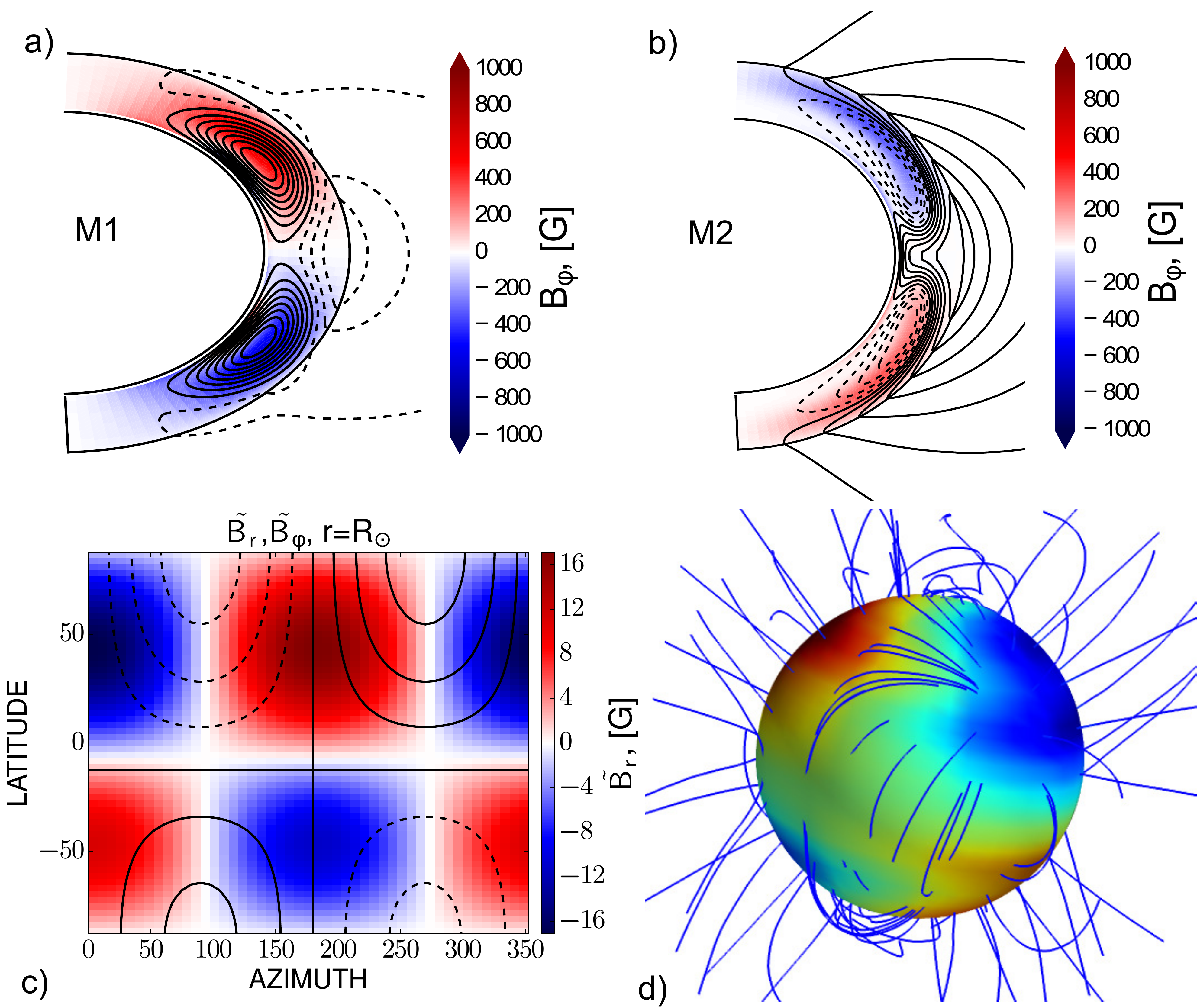}

\protect\protect\protect\caption{\label{fig:Initial-nonaxisymmetric-field}Axisymmetric and non-axisymmetric
field structure at the moment of initialization of the non-axisymmetric
perturbation: a) distribution of the axisymmetric toroidal magnetic
field (background image) and poloidal field lines in the meridional
cross-section for model M1; b) the same as a) for model M2; c) components
of the non-axisymmetric magnetic field at the surface, $\tilde{B}_{r}$
is shown as color image, $\tilde{B}_{\phi}$ is shown by contour lines
plotted every 3G; d) illustration of the magnetic field lines in model
M1 at the moment of initialization of the perturbation.}
\end{figure}

{We simulate the sunspot number $W$ using the anzatz \citep{pipea2012AA}:
\begin{equation}
W\left(t\right)=\left\langle B_{{\rm max}}\right\rangle \exp\left(-\frac{B_{0}}{\left\langle B_{{\rm max}}\right\rangle }\right),\label{eq:wolf}
\end{equation}
where in model M1 $\left\langle B_{{\rm max}}\right\rangle $ is the
maximum strength of the toroidal magnetic field averaged in the subsurface
layers over radius in the range of $0.9-0.99R_{\odot}$, and $B_{0}$
is a characteristic strength of the toroidal magnetic field, $B_{0}=800$G.
In model M2 we measure $\left\langle B_{{\rm max}}\right\rangle $
in the layer of convergence of the two meridional circulation cells,
which is in the range of $0.85-0.9R_{\odot}$. This layer corresponds
to the maximum of the toroidal magnetic field strength in the convection
zone for model M2. The AS model with the double-cell meridional circulation
was discussed in details in our previous papers \citep{PK13,2014ApJ_pipk}.}

\subsection{Initial conditions}

In our first runs the weak initial field, which consisted of a superposition
sum of polar and equatorial dipoles with the magnetic field strength
of $0.01$G, evolved to a state in which the axisymmetric dynamo regime
dominates. In this regime, the typical strength of the axisymmetric
toroidal field in the convection zone is about 1kG. However, the non-axisymmetric
field is rather weak with the strength about 10$^{-5}$G. This means
that in our model the non-axisymmetric magnetic field is linearly
stable, unless the dynamo governing parameters are forced to be non-axisymmetric,
e.g., like in the paper by \citet{ruz04}.

Exploring the nonlinear solutions we found that the evolution of the
non-axisymmetric field depends on the initial conditions which include
the strength and geometry of both the axisymmetric $\mathbf{\overline{B}}$-field
and non-axisymmetric $\mathbf{\tilde{\mathbf{B}}}$-field. The evolution
of the large-scale magnetic field depends on the presence of the meridional
circulation, too.

In the following section we present results for the non-axisymmetric
dynamo which was perturbed by a finite-amplitude non-axisymmetric
$\mathbf{\tilde{\mathbf{B}}}$-field in the developed axisymmetric
dynamo regime. Such non-axisymmetric perturbations can be developed
either due to evolution of active regions or due to instabilities
not described by the mean-field theory (e.g., \citealt{2001ApJ...559..428D}).
For the seed field we consider a non-symmetric relative to the equator
perturbation represented by a sum of the equatorial dipole (l=1, m=$\pm$1)
and quadrupole ($l=2$, $m=\pm1$) components. In Eq(\ref{eq:sdec})
we define 
\begin{align}
S_{1,1} & =\frac{1}{2}\left(1-\mathrm{erf}\left(\frac{\left(r_{s}-r\right)}{d}\right)\right)\frac{r_{e}}{r},\label{eq:in1}\\
S_{1,2} & =\frac{1}{2}\left(1-\mathrm{erf}\left(\frac{\left(r_{s}-r\right)}{d}\right)\right)\left(\frac{r_{e}}{r}\right)^{2},\label{eq:in2}
\end{align}
where, $r_{e}=0.99R_{\odot}$, $r_{s}=0.9R_{\odot}$ is the bottom
of the subsurface shear layer, $d=0.02R_{\odot}$. The other $S_{l,m}$
and $T_{l,m}$ coefficients are zero in the perturbation. The initial
non-axisymmetric perturbation is concentrated in the near-surface
shear layer. The depth of the non-axisymmetric perturbation can influence
the evolution of the axisymmetric dynamo. {At the moment of
the initialization of perturbation the strength of the axisymmetric
toroidal field is by two orders of magnitude greater than of the non-axisymmetric
one. At the same time the maximum amplitude of the NA radial magnetic
field at the surface is about twice of the axisymmetric one (see Fig.
2c).} {Also, the initialization time we have the following
parameters for the parity of the AS and NA parts are $\overline{P}=-1$,
$\tilde{P}=-0.3$ respectively and the indexes of the axisymmetry
are $M\approx10^{-4}$. }

\begin{table}
\protect\protect\protect\caption{Summary of the dynamo models and their parameters.}

\begin{tabular}{|c|>{\raggedright}p{0.3\textwidth}|>{\raggedright}p{0.3\textwidth}|}
\hline 
Components  & M1  & M2\tabularnewline
\hline 
\hline 
The $\boldsymbol{\mathcal{E}}$  & Eq.(\ref{eq:EMF-1})  & same\tabularnewline
\hline 
circulation  & no  & $\overline{U}=10$m/s\tabularnewline
\hline 
free parameters, (see,\citealt{pi15M})  & $C_{\alpha}=0.04$,$C_{\delta}=C_{\alpha}/3$, $C_{\eta}=0.06$, a=3,
$C_{\beta}=1$ and $C_{\beta}=C_{\eta}$, $R_{m}=10^{4-6}$, $\alpha_{MLT}=2$  & $C_{\alpha}=0.05$, others are same as in M1\tabularnewline
\hline 
\end{tabular}
\end{table}

\section{Results}

Figure \ref{fig:Initial-nonaxisymmetric-field} illustrates the geometry
of the axisymmetric and non-axisymmetric fields just before initialization
of the non-axisymmetric perturbation. {The perturbation is
initialized at $t\approx13.2$ years for model M1 and $t\approx5.7$
for model M2.} {Those times corresponds approximately the same
phase of the AS toroidal magnetic field in the upper half of the convection
zone in the models.} Figure \ref{fig:Time-latitude-diagram-(a),}
illustrates the evolution of the axisymmetric magnetic field before
and after the perturbation for two models, with and without the meridional
circulation. We show the time-latitude diagrams for the toroidal magnetic
field in the subsurface shear layer and the radial magnetic field
at the surface. In the model with the meridional circulation the toroidal
magnetic field is shown for the middle of the convection zone (see
\citealt{PK13}). The radial evolution is shown for $30^{\circ}$
latitude in the Northern hemisphere. 

\begin{figure}
\includegraphics[width=1\textwidth]{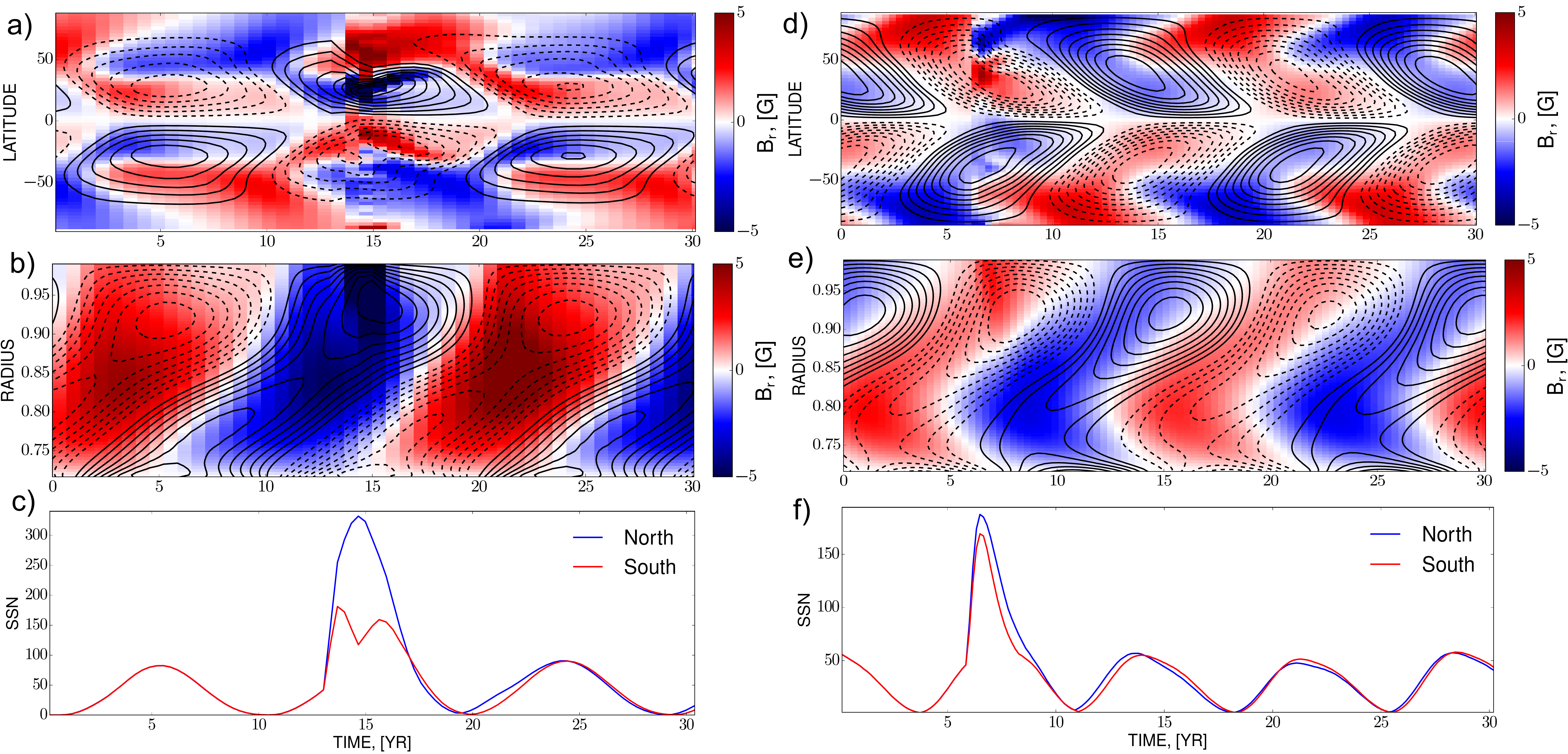}

\protect\protect\protect\caption{\label{fig:Time-latitude-diagram-(a),}. a-c) Dynamo evolution in
model M1 before and after the initialization of the non-axisymmetric
perturbation at $t=13$ yr: a) time-latitude diagram, b) time-radius
diagram at 30$^{\circ}$ latitude, c) the simulated sunspot number
for the Northern and Southern hemispheres; d-f) the same as in a-c)
for model M2. Color images show the radial magnetic field component.
The contour lines show the toroidal component, plotted every $100$
G.}
\end{figure}

The models show that the imposed non-axisymmetric perturbation produces
a transient cycle in both models M1 and in M2. In the Northern hemisphere,
where the initiated perturbation is greater, the simulated sunspot
number cycle is stronger than in the Southern hemisphere. The perturbation
affects the reversal of the polar magnetic fields. In the Northern
hemisphere the polar field reversal occurs earlier than in the Southern
hemisphere where we see multiple reversals. Model M1 shows a time
shift of about 2 years for the polar reversals in the Northern and
Southern hemispheres. We find that these phenomena depend on the depth
of perturbation (parameters, $r_{s}$ and $d$ in Eq(\ref{eq:in1})).
For instance, the polar field reversal happens earlier in the Southern
hemisphere than in the Northern one if the imposed perturbation is
shallower $r_{s}=0.95R_{\odot}$. We also see that in both models
the cycle returns quickly to the previously established axisymmetric
state. {Additional runs which are not illustrated here show
that this restoration could take a longer time interval (more than
1 cycle) if the axisymmetric field would have a mixed parity at the
initialization moment. }

\begin{figure}
\includegraphics[width=0.9\textwidth]{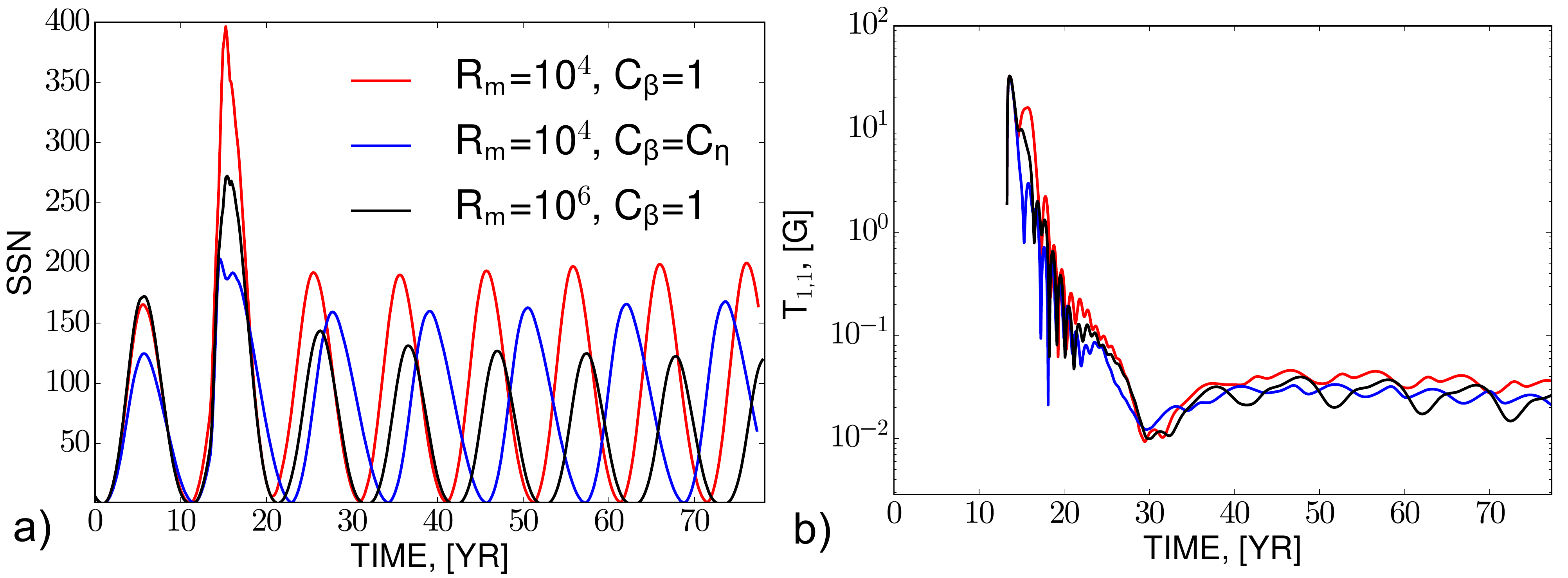}

\protect\protect\protect\caption{\label{fig:EvR}Evolution of the simulated sunspot number (a) and
$T_{1,1}$ harmonic (b) in model M1 for the different values of the
parameter $R_{m}$ and $C_{\beta}$: $R_{m}=10^{4}$, $C_{\beta}=1$
(red), $R_{m}=10^{4}$, $C_{\beta}=C_{\eta}$ (blue), $R_{m}=10^{6}$,
$C_{\beta}=1$ (black).}
\end{figure}

The evolution of the magnetic helicity density (Eq \ref{eq:helcon-1})
depends on the magnetic Reynods number, $R_{m}$, which is a free
parameter of the model. For higher $R_{m}$ the relaxation in both
models is similar. Figure \ref{fig:EvR} illustrates the results for
model M1 for $R_{m}=10^{4}$ and $R_{m}=10^{6}$ and also for model
M1 with a reduced magnetic buoyancy effect ($C_{\beta}=C_{\eta}$).
The developed axisymmetric dynamo regime which is employed at the
beginning of evolution series shown in Figure \ref{fig:EvR} is related
to the the case $R_{m}=10^{4}$ and $C_{\beta}=1$. This explains
why the magnetic cycles in model M1 with $C_{\beta}=C_{\eta}$ and
$R_{m}=10^{6}$ do not relax to the original cycle amplitude. The
model with a reduced magnetic buoyancy shows a smaller (by a factor
2) magnitude of $T_{1,1}$ mode after the relaxation. The mode $T_{1,1}$
shows a larger variations of amplitude in case of $R_{m}=10^{6}$
than in the case of $R_{m}=10^{4}$ after the relaxation.

\begin{figure}
\includegraphics[width=0.7\textwidth]{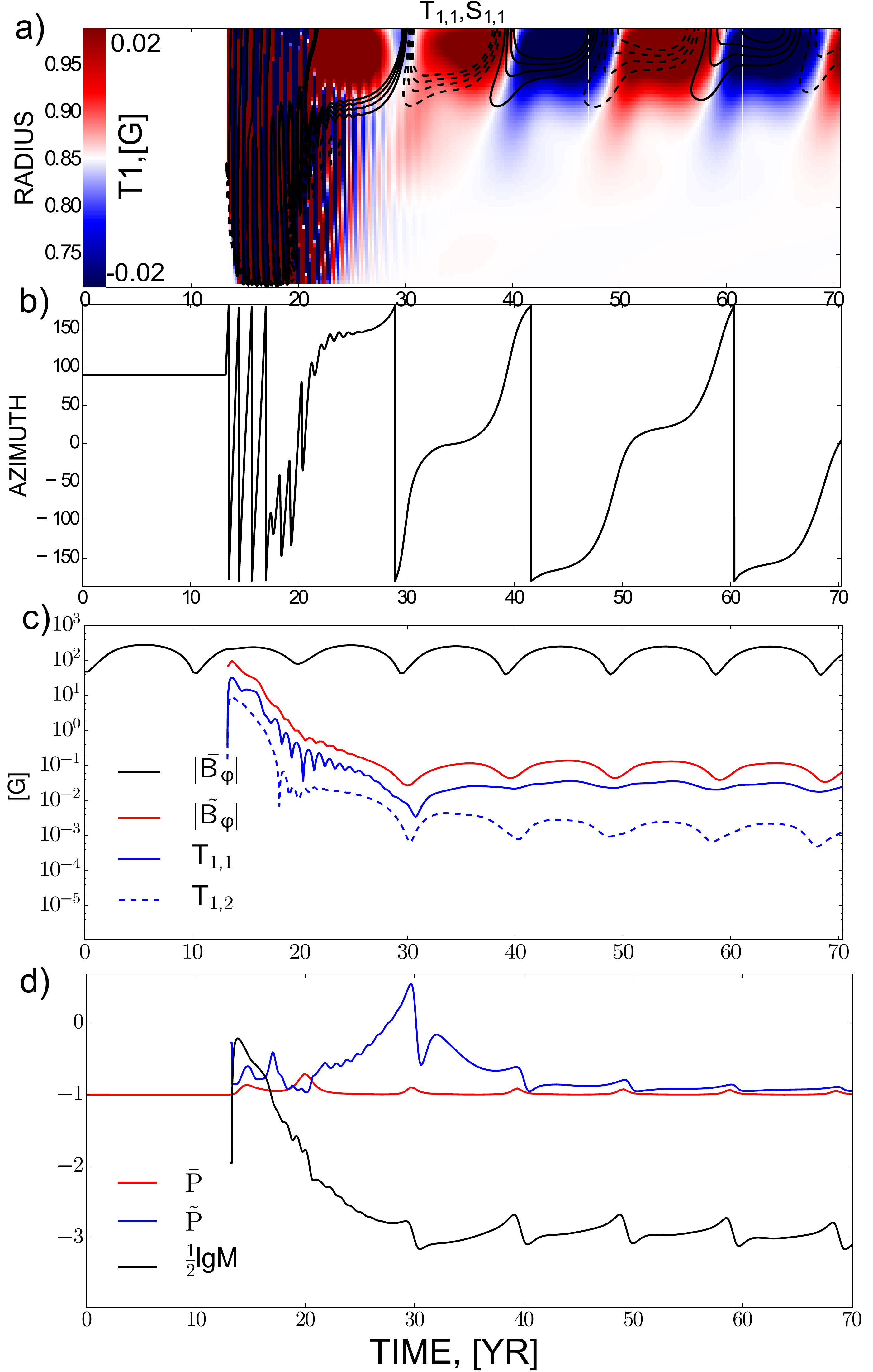}

\protect\protect\protect\caption{\label{fig:Evolution}Non-axisymmetric modes of model M1: a) the time-radius
evolution of the dynamo modes $T_{1,1}$ (background image) and $S_{1,1}$
modes (contours are in the same range of values as the color scale);
b) evolution of longitude of the maximum of the large-scale toroidal
field, c) the mean strength of the near-surface axisymmetric and non-axisymmetric
toroidal magnetic field and the strength of the m=1 T-potentials,
d) evolution of parities, $\overline{P}$ and $\tilde{P}$, and the
index of the axisymmetry $M$. }
\end{figure}

\begin{figure}
\includegraphics[width=0.7\textwidth]{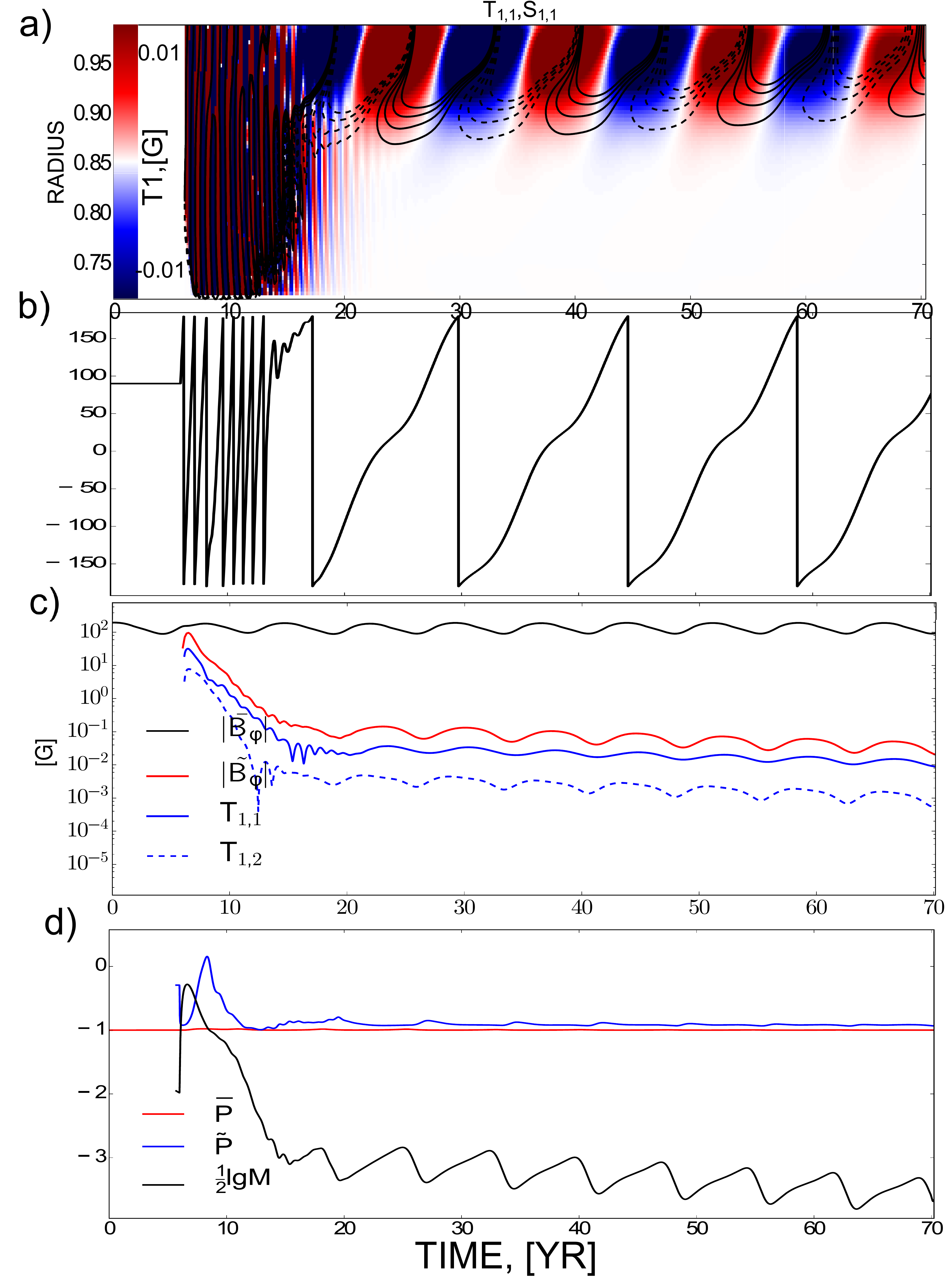}

\protect\protect\protect\caption{\label{fig:Evolution-1}The same as Figure \ref{fig:Evolution} for
model M2}
\end{figure}

\begin{figure}
\includegraphics[width=0.9\textwidth]{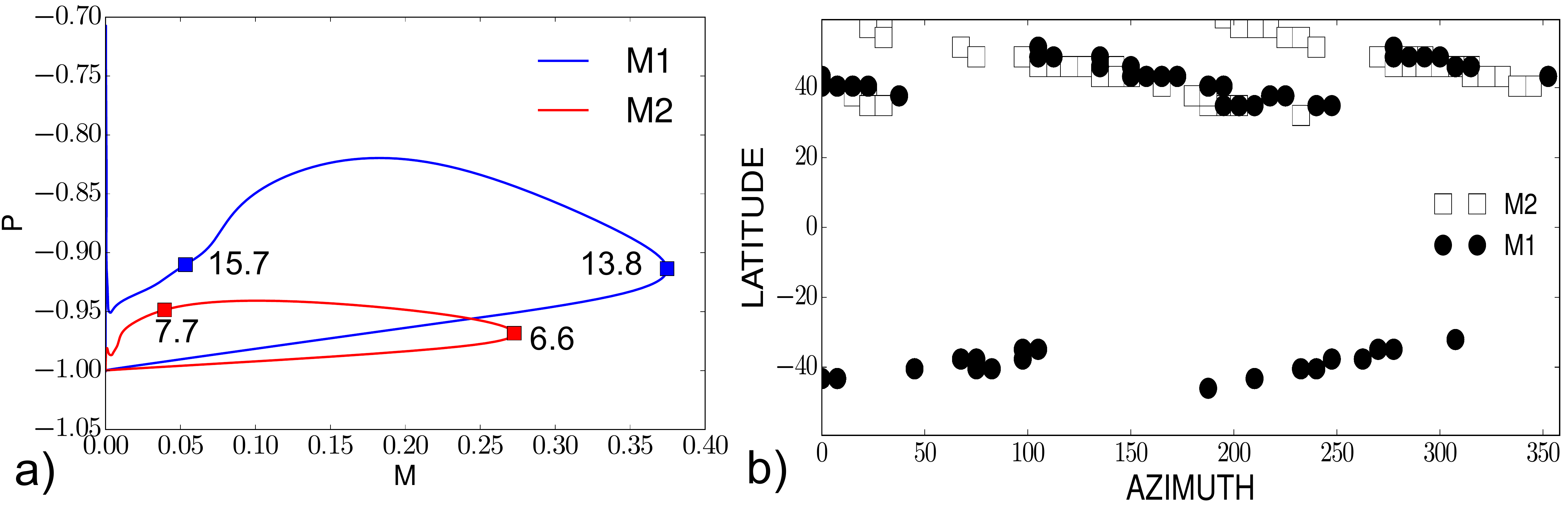}

\protect\protect\protect\caption{\label{fig:MP} a) Evolution of parity of total magnetic field $P$
versus the axisymmetry index $M$ for model M1 (blue) and M2(red);
b) Position of maxima of the total toroidal magnetic field strength
of the subsurface shear layer (r=0.92R) in models M1 and M2 after
relaxation.}
\end{figure}

The restoration of the initial of the axisymmetric magnetic field
evolution after the perturbation does not mean that the non-axisymmetric
field completely dissipates. {Figures \ref{fig:Evolution},
\ref{fig:Evolution-1}(a,c,d) show that a low strength non-axisymmetric
field is maintained in the model. We find that the strength of the
toroidal field T-potentials is reduced from about 100G at the maximum
to 0.01G after the relaxation. In model M2 the NA magnetic field decays
slowly after relaxation. Both the M1 and M2 models show a deviation
of the parity of the AS magnetic field, $\overline{P}$ from -1 (corresponding
to antisymmetric about theequator magnetic field) after the perturbation.
The NA magnetic field shows a mixed parity solution during and after
the relaxation phase. The important finding is that the NA perturbation
evolves through a growing phase for about half an year in model M1
and for about one year in model M2, and it is not just decaying after
the initialization. Variations of the axial symmetry index, $M$,
and the parity, $P$, is shown in Fig\ref{fig:MP}(a). We see that
$M$ first grows and then after about 1 year the perturbation is reflected
in the parity of the large-scale magnetic field. Model M1 shows greater
variations of the parities than the model M2. We find that in the
M1 relaxation of the parity takes more than one cycle. After relaxation
the dynamo model returns to axial symmetry, $M\approx10^{-6}$.} Even
such a low strength non-axisymmetric magnetic field can produce some
interesting phenomena which may be related to solar observations.

Figures \ref{fig:Evolution},\ref{fig:Evolution-1}(b) show the longitudinal
evolution of a maximum of the large-scale toroidal field in the subsurface
shear layer. {The longitude is computed for the coordinate
frame rotating with the period of about 25 days. After the perturbation
initialization the longitude of the m=1 mode drifts around the Sun.
There is no fluid motion associated with this drift, this drift is
an analogue of the latitudinal dynamo waves in the slowly-rotating
regime. The effect was suggested earlier in the study of the linear
dynamo regimes by \citet{rad86AN}. It was found in observations (e.g.,
\citealt{tuom02,lin2013}) and in the direct numerical
simulations \citep{col14}. The speed of the drift changes after the
relaxation. Model M1 shows the almost fixed positions for the longitude
of the m=1 mode during epochs of the maximum of the AS toroidal magnetic
field. }The latitude-longitude position of the maxima of the large-scale
toroidal magnetic field strength after relaxation are illustrated
in Fig\ref{fig:MP}(b). Model M1 (without the meridional circulation)
shows the periodic changes of the longitude by 180$^{\circ}$ degrees
during the magnetic cycle decay. The change of the longitude is accompanied
by a change of the hemispheric position of the field maximum. Thus,
the orientation of the global non-axisymmetric field is reversed every
cycle during the minima of the toroidal magnetic field . This behavior
may correspond to the ``flip-flop'' phenomenon of the active longitudes
suggested for stellar magnetic cycles \citep{ber2004}.

Model M2 which includes the meridional circulation has no stable positions
of the non-axisymmetric magnetic field azimuth. Figures \ref{fig:Evolution}(a)
and \ref{fig:Evolution-1}(a) clarify the reason for this. The effect
disappears because the circulation mixes the magnetic field in the
subsurface shear layer with the magnetic field of the deep interior
with a period which approximately corresponds to the period of the
magnetic cycle. We see that oscillations of the $S_{1,1}$ and $T_{1,1}$
harmonics have a $\pi/2$ phase shift. Thus, for a persistent appearance
of the active longitude the phases of the $S_{1,1}$ and $T_{1,1}$
harmonics should be consistent. Model M2 shows a continuous drift
of the longitude of the large-scale toroidal magnetic field strength
in the course of the magnetic activity evolution.{ Additional
runs which are not illustrated here show that the model with the circulation
could produce a sort of active longitude phenomenon but for another
combination of the circulation cells differen from the solar case
shown in Figure\ref{one}(b). More specifically, the strength of the
bottom cell should be reduced compared to model M2. In this case we
find that the active longitudes occupy only the Northern hemisphere.}

\begin{figure}
\includegraphics[width=1\textwidth]{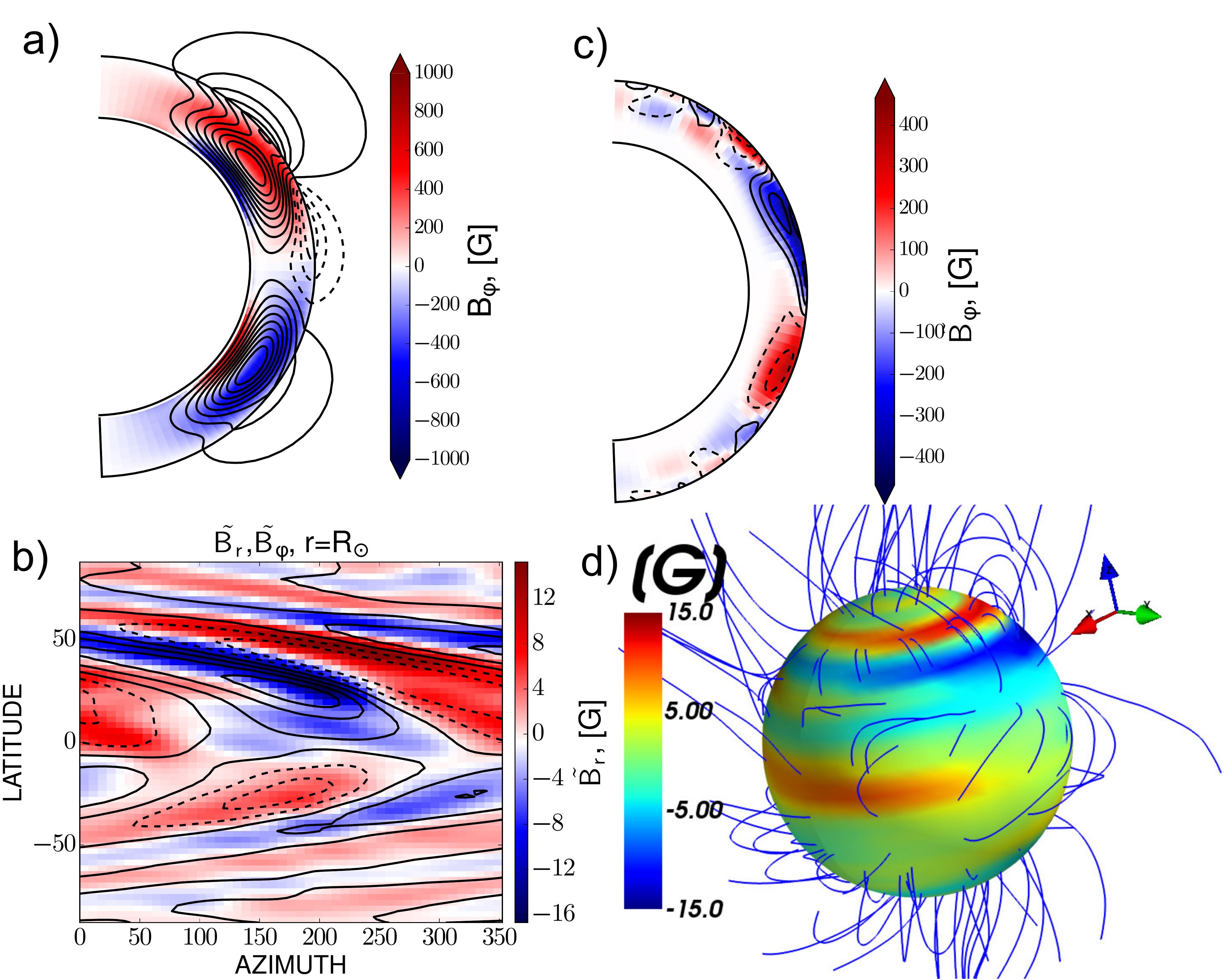}

\protect\protect\protect\caption{\label{fig:1-year-after}Model M1 a half year after the initialization
of the non-axisymmetric perturbation, a) distribution of the axisymmetric
toroidal magnetic field (background image) and poloidal field lines
in the meridional cross-section, b) azimuth-latitude distribution
of the non-axisymmetric radial magnetic field (background image) and
toroidal magnetic field (contours are in the same range as the color-scale)
on the surface; c) the same as in panel (a) for the non-axisymmetric
components of magnetic field; d) the magnetic field lines of the non-axisymmetric
magnetic field above the surface.}
\end{figure}

\begin{figure}
\includegraphics[width=1\textwidth]{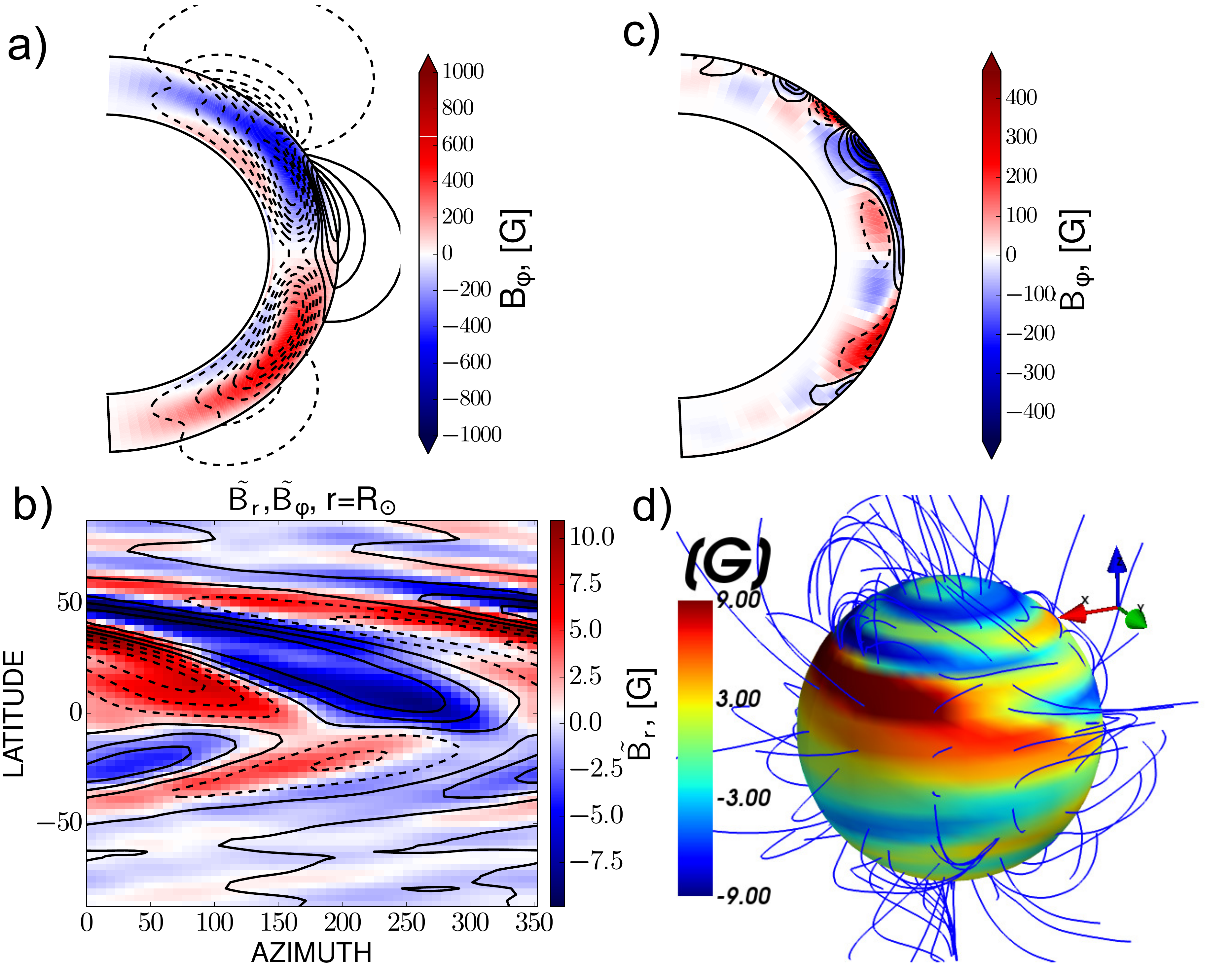}

\protect\protect\protect\caption{\label{fig:1-year-after-1}The same as Figure \ref{fig:1-year-after}
for model M2}
\end{figure}

Figure \ref{fig:1-year-after} shows snapshots of the axisymmetric
and non-axisymmetric fields after a half-year evolution of the initial
perturbation. We also show the configuration of the external potential
magnetic field. This period of time corresponds to a maximum of the
toroidal magnetic field in the upper part of the convection zone,
and the epoch of the polar field reversal. The non-axisymmetric part
of the field is concentrated to the surface (as the initial field).
The longitude-latitude diagram shows the distributions of the large-scale
non-axisymmetric magnetic field. It illustrates how the differential
rotation stretches the initial magnetic field configuration (cf.,
Fig\ref{fig:Initial-nonaxisymmetric-field}a). The snapshots in Figures
8 and 9 show the large-scale unipolar regions which extend from the
equator to the high-latitude regions the poles. The increasing of
the non-axisymmetric magnetic field in the polar regions results in
twisted field lines in the polar caps.

\begin{figure}
\includegraphics[width=1\textwidth]{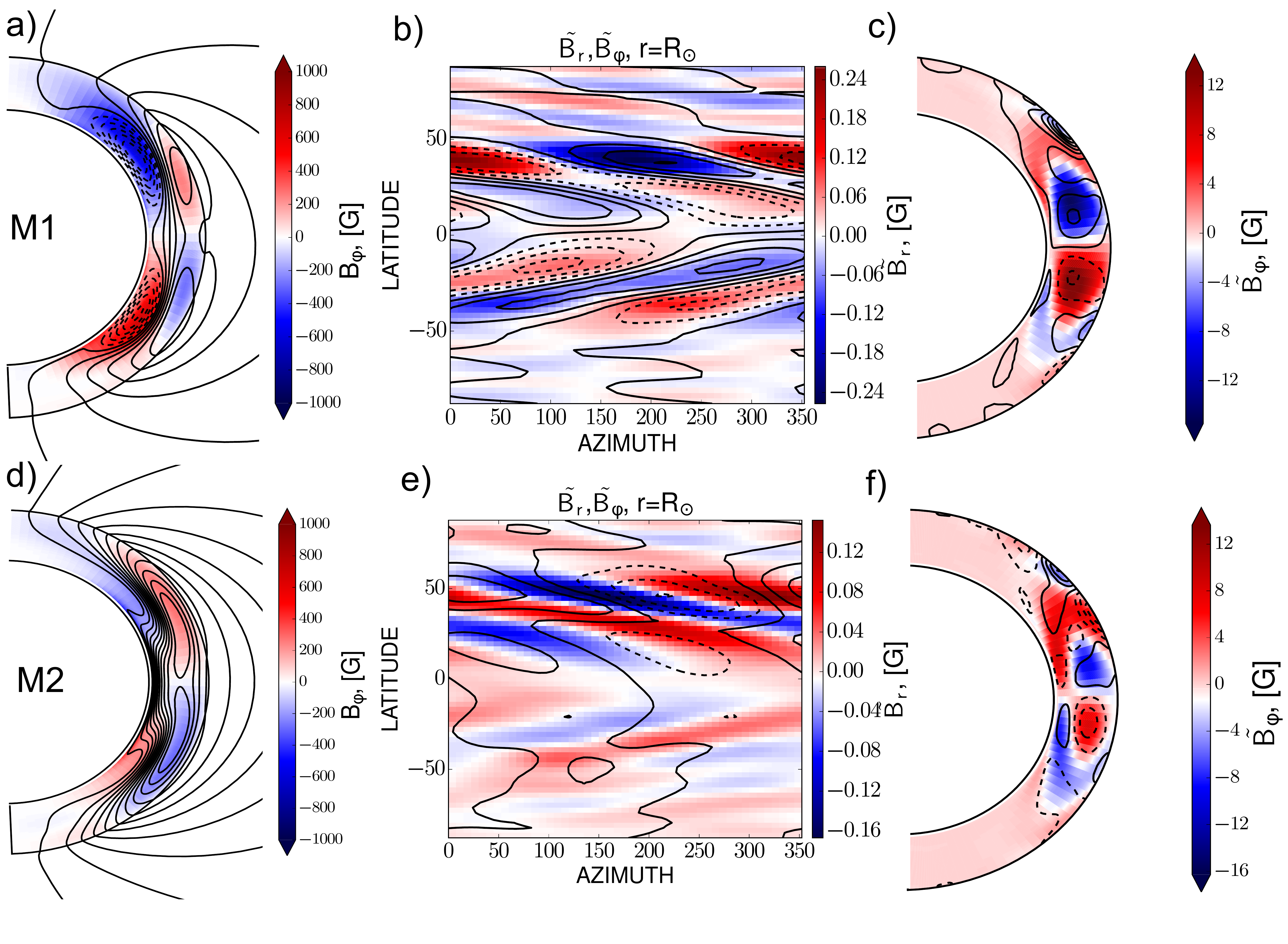}

\protect\protect\protect\caption{\label{fig:The-same-as}The same as in Fig. 7 after 6 years of the
evolution after the initialization of the non-axisymmetric perturbation.}
\end{figure}

Figure \ref{fig:The-same-as} illustrates snapshots of the magnetic
field configuration during the decaying phase of the magnetic cycle
in model M1 after 6 years from the initialization. We see that the
non-axisymmetric toroidal field distributed over the convection zone
has maxims at the bottom of the convection zones and in the near equatorial
region. The strength of the non-axisymmetric field is much smaller
than the strength of the axisymmetric field. Model M2 has long overlaps
between the subsequent cycles. Snapshots for this model are shown
in Fig.\ref{fig:The-same-as}(bottom) for the growing phase of the
cycle. The snapshots show the situation when the symmetric with respect
to the equator m=1 mode dominates at the surface. In the deep layers
the general distribution of the non-axisymmetric magnetic field is
close to model M1.

The stationary dynamo evolution begins about 15 years after the initialization
of the non-axisymmetric magnetic field in model M1 (see Fig.\ref{fig:Evolution}).
The relaxation time of model M2 is about one cycle. In the stationary
stage the non-axisymmetric field is concentrated at the top of the
convection zone like in the snapshots shown in Figures \ref{fig:1-year-after}
and \ref{fig:1-year-after-1} (also see Figures \ref{fig:Evolution}(a)
and \ref{fig:Evolution-1}(a)). The antisymmetric m=1 mode with mixed
parity dominates in both models.

{In addition to models M1 and M2 presented in the paper, we
calculated the models for different initial conditions by changing
the spatial distribution of the non-axisymmetric perturbation and
the initialization time relative to the different epochs of the magnetic
cycle. In model M1 the effect of perturbation is the greatest when
it is initiated at the growing phase of the cycle. Also, the impact
of the perturbation, and the amplitude of the non-axisymmetric field
after the relaxation are stronger with the increase of the perturbation
depth. However, if the perturbation in the form of Eq(\ref{eq:in1})
is located near the bottom of the convection zone, it produces only
a weak effect on the large-scale distributed dynamo.}

\begin{figure}
\includegraphics[width=1\textwidth]{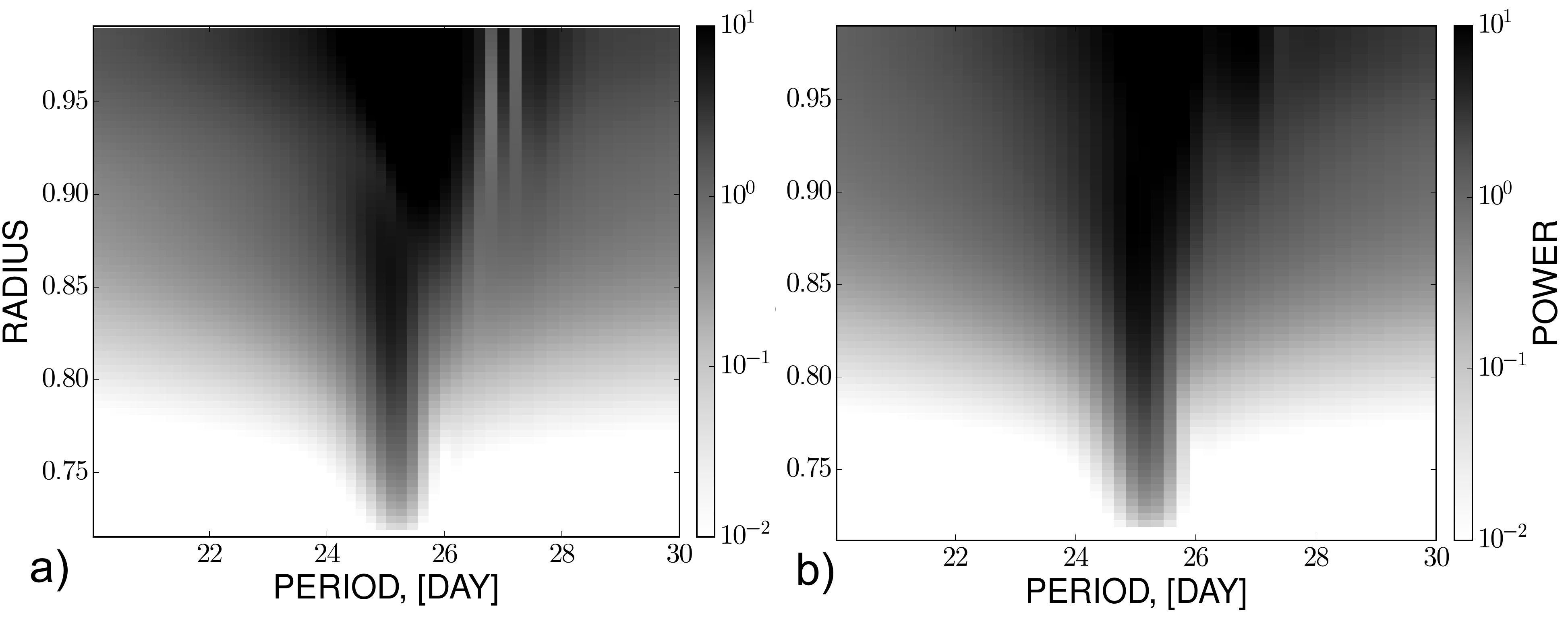}\protect\protect \protect\caption{\label{spec} The power spectra for rotation periods of azimuth of
non-axisymmetric mode $S_{1,1}$ in the solar convection zone for:
a) model M1; b) model M2.}
\end{figure}

To investigate the rotation rate of the non-axisymmetric modes we
calculate power spectra of the $m=1$ mode azimuth for different radii.
Figure \ref{spec} shows results for the three levels of the solar
convection zone. The time series cover the period of about 5 year
after the initialization of the non-axisymmetric field in the models.
At the initialization period the equatorial dipole was rotating with
a period of 27.26 days, which corresponds to the differential rotation
period at the latitude of the maxima of the initial perturbation.
For model M1 it is found that in the subsurface shear layer the equatorial
dipole rotates with the periods of 25.1-25.5 days. This corresponds
to rotation of the subsurface shear layer at 30$^{\circ}$ latitude.
At the surface the dipole rotates with the period about 25.7 days,
and at the bottom of the convection zone it rotates with the period
of 25.1 days. The origin of these rotational periods has to be studied
further. It seems that the periods of rotation follow the rotation
profile in the solar convection zone, e.g., see the typical bow of
the magnetic field distribution in Fig \ref{fig:The-same-as}(c).
In model M2 the meridional circulation mixes all layers of the convection
zone, producing a unique maximum of the non-axisymmetric magnetic
field rotating with period of 25.4 days, which corresponds to the
differential rotation period at latitude of 30$^{\circ}$ and $r=0.9R$.

Finally, the animated evolution of the large-scale magnetic field
is illustrated by two videos:
``http://www.youtube.com/watch?v=0i5tslwaxao'' and \\
''http://www.youtube.com/watch?v=buNK91Sb3OA'', showing results for model
M1.

\section{Discussion and Summary }

In the paper we explored the evolution of a non-axisymmetric (NA)
magnetic field perturbations in the mean-field solar-type dynamo models.
The models are kinematic with respect to the mean flow. The distribution
of the mean flow is taken from the recent results of helioseismology,
including the subsurface rotational shear layer and the double-cell
meridional circulation pattern which was suggested recently by results
of \citet{Zhao13m}. We studied models with and without meridional
circulation. The non-axisymmetric dynamo model takes into account
the mean turbulent electromotive force in a fairly complete form.
The mean electromotive force which is employed in the non-axisymmetric
part of the model is the same as for the axisymmetric (AS) part except
that the $\delta$-effect ($\Omega\times J$ term) \citep{rad69}
was omitted in the non-axisymmetric electromotive force. We plan to
investigate it separately.

{The dynamo models studied in the paper show two distinct evolution
phases. The transient phase starts after the initialization of the
NA perturbation and ends approximately after one dynamo cycle period.
The results suggest that the non-axisymmetric magnetic field can considerably
affect the axisymmetric dynamo. This effect depends on the amplitude
of the perturbation, the depth of perturbation, and the phase of evolution
of the axisymmetric magnetic field. In the paper we illustrated the
effect of a perturbation representedby the mix of the odd and even
parities of $m=1$ mode of magnetic field with magnitude of 1G. The
depth of the perturbation is $r=0.9R_{\odot}$ . }

{The models show that during the transient phase the NA magnetic
field is amplified, to the level of the axysymmetry index $M=0.3$,
which is 30\% of the total magnetic field energy. It affects the North-South
asymmetry of the magnetic field about equator. The growth and decay
of the NA magnetic field is accompanied by oscillations with a period
about of 1.5 year. Rotation of the$m=1$mode during the transient
phase shows a continuous spectrum of the rotation periods because
the evolution of the NA magnetic field is strongly coupled with the
differential rotation. The maximum in the spectrum of the rotation
periods is at about of 25.4 days.}

{The transient phase of the dynamo evolution follows by a ``stationary''
phase when the NA magnetic field evolves slowly, varying on the time
interval which is longer than the dynamo period. Solutions for the
second phase could be compared with results of the previous NA dynamo
models (see, \citealt{rad86AN,radler90,moss91,moss99,el2005}). In
the ``stationary'' phase we found only a weak NA magnetic field
($M\sim10^{-6}$) with the dominant $m=1$ structure and with the
antisymmetric relative to the equator magnetic parity. The NA dynamo
modes rotates rigidly. }

{Model M1 (without the meridional circulation) shows that the
rigid rotation of the m=1 dynamo modes is accompanied by changes of
the longitude and the hemispheric position of the maximum of the $m=1$
mode with a period of one dynamo cycle. This regime resembles the
so-called ``flip-flop'' phenomenon which is found in the stellar
magnetic activity \citep{jet1991,kon2002,ber2004,lin2013}.
It was also previously suggested by other NA dynamo models \citep{tuom02,moss04,el2005,berd06}.
The flip-flop phenomenon may be related another phenomenon so-called
the ``active longitudes'' (AL) (e.g.,\citealt{vitinsk,vetal86}).
In Introduction it was mentioned that the persistence the AL on the
Sun on the century time scale remains a highly controversial issue.
The origin of flip-flop phenomenon in nonlinear non-axisymmetric dynamo
models was discussed in details by \citet{moss04} (see, also, discussion
in \citealt{berd06}). In our calculations the active longitude is
fixed when the $T_{1,1}$ and $S_{1,1}$ dynamo-modes are in phase.
The flip-flop occurs when the orientation of the equatorial dipole
changes the sign. If this effect is accompanied by the equatorial
symmetry variations of the axisymmetric magnetic field then the orientation
of the large-scale magnetic field changes by 180$^{\circ}$. The current
explanation remains qualitative because the amplitude of the NA magnetic
field in the model is rather small (the axial symmetry index, $M\sim10^{-6}$).
It is not clear how such a weak NA magnetic field could modulate the
sunspot activity. }

{One interesting feature of model M1 is that in the stationary
evolution phase the antisymmetric parity of the $m=1$ mode dominates.
This is different from previous results of the linear theory by \citet{rad86AN}
and the previous nonlinear models \citep{radler90,moss91,moss99}.
Also the maximum of the $m=1$ is located on the mid-latitude zone
which is different from the typical poleward concentration of the
NA magnetic field suggested by results of the above cited papers.
The origin of this disagreement is unclear. We have to stress that
in our models we employ a fairly complete information about the radial
profiles of the $\alpha$-effect and the differential rotation suggested
by the modern results of the helioseismology, and results of recent
theoretical works. Also we include the magnetic helicity conservation
which was not taking into account before in the non-axisymmetric dynamo
models. The dynamo parameters such as the $\alpha$-effect and the
turbulent diffusion coefficients were tuned to reproduce the 22 year
dynamo cycle by the axisymmetric dynamo. Considering the ratio between
the $\alpha$ and $\Omega$-effects in our models, we find ${\displaystyle \frac{\Omega_{0}R_{\odot}}{\alpha_{0}}\sim10^{3}}$,
where $\alpha_{0}$ is the magnitude of the $\alpha$-effect. This
is about by 2 orders of magnitude than the quantity employed by \citet{radler90}
and \citet{moss99}. The finding similar to ours, i.e., a concentrated
to the equator antisymmetric $m=1$ magnetic field mode was reported
by \citet{2013ApJ...762...73N} for the global numerical simulation
of the convective dynamo on the solar-type star rotating 3 times faster
the modern Sun.}

{Model M2 with the meridional circulation illustrates another
feature predicted by the linear analysis of the kinematic NA dynamo
models. In this model the longitudinal position of the $m=1$ mode
in the stationary phase of evolution drifts around the Sun with a
period of about one dynamo cycle. This can be interpreted as an azimuthal
dynamo wave. The effect was predicted by the mean-field theory (see,
e.g. \citealt{KR80}), it was found in the mean-field NA dynamo models
\citep{rad86AN,radler90,moss91}, and in the direct numerical simulations
\citep{col14}. It was also found in observations of magnetic activity
on fast-rotating late-type stars (e.g., \citealt{tuom02,lin2013}).}

\section{Conclusions}

We considered non-axisymmetric mean-field dynamo models including
the non-linear magnetic helicity and magnetic buoyancy effects, which
were not studied before. The study confirms the previous findings
of \citet{moss99} that the non-axisymmetric dynamo component is rather
weak if we start from a weak initial (seed) non-axisymmetric field.
We notice that our models can be characterized as weakly non-linear
because the parameter of the $\alpha$-effect in the model is only
30\% above the dynamo instability threshold. Also the magnetic helicity
conservation and magnetic buoyancy prevent the generation of magnetic
field of the super-equipartition strength \citep{2007NJPh....9..305B,hub-br12}.
Thus, the low-strength non-axisymmetric magnetic field generated from
a weak seed field can be explained by the linear stability of the
non-axisymmetric field, and by the weak non-linearity of the dynamo
system. However, finite-amplitude non-axisymmetric perturbations,
which can be developed in the complex dynamical system may have significant
effects on the dynamo process.

The modeling results show that the magnetic helicity conservation
(also known as dynamical quenching of the $\alpha$-effect) is an
important factor to preserve the non-axisymmetric field from a complete
decay. It is found that for the magnetic Reynolds number $R_{m}=10^{6}$
the coupling of the non-axisymmetric magnetic field with the axisymmetric
dynamo process is stronger than in case of $R_{m}=10^{4}$. The strong
coupling results in a synchronization between oscillations of the
non-axisymmetric and axisymmetric magnetic fields. Also, our models
include nonlinear effects of magnetic buoyancy. It is found that if
the magnetic buoyancy effect is switched off then the strength of
the non-axisymmetric field after the relaxation is decreased by a
factor of two.

The paper illustrates our initial results of the nonlinear non-axisymmetric
mean-field dynamo model. The axisymmetric part of the model is based
on our previous results. The non-axisymmetric perturbations are assumed
to be located in the near-surface rotational shear layer. For the
first time we demonstrate that nonlinear coupling between the asymmetric
and the non-axisymmetric fields can impact the generation of the axisymmetric
field in the case of finite-amplitude perturbations. The effect depends
strongly on the dynamo mechanisms involved in the problem, the spatial
distribution of perturbation, the phase of the dynamo cycle at the
time of initialization of the perturbation, and on how well the magnetic
helicity is conserved in the system. These factors determine the subsequent
evolution of the dynamo system including the dynamo cycle, evolution
and rotation of the non-axisymmetric modes of large-scale magnetic
fields.

In summary: 
\begin{itemize}
\item {The differential rotation and interaction with the axisymmetric
magnetic fields results to amplification of the non-axisymmetric perturbation
during the transient phase of evolution, producing the magnetic field
configuration which deviates considerably from the axial symmetry. }
\item {In the solar dynamo models non-axisymmetric magnetic field
perturbations developed in the near-surface rotational shear layer
affect the strength and period of the dynamo cycles. }
\item {Without the meridional circulation the non-axisymmetric dynamo-mode
shows a ``flip-flop'' phenomenon and also has a fixed longitudinal
position resembling the active longitude phenomenon. However, the
solar-type (double-cell) meridional circulation destroys these effects.
Instead, the non-axisymmetric field represent a travelling in azimutal
direction dynamo wave. }
\end{itemize}
Acknowledgments The work was partially supported by NASA grants NNX09AJ85G
and NNX14AB70G. VP thanks support of RFBR under grants 14-02-90424,
15-02-01407 and the project II.16.3.1 of ISTP SB RAS.


\end{document}